# OpenTripPlanner, OpenStreetMap, General Transit Feed Specification: Tools for Disaster Relief and Recovery

Chelcie G. Narboneta
Department of Information Systems and Computer Science
Ateneo de Manila University
Loyola Heights, Quezon City 1108, Philippines
cgn1108@yahoo.com

Kardi Teknomo
Department of Information Systems and Computer Science
Ateneo de Manila University
Loyola Heights, Quezon City 1108, Philippines
teknomo@gmail.com

*Abstract*— OpenTripPlanner was identified as the most promising open source multimodal trip planning software. OpenStreetMap, which provides mapping data to OpenTripPlanner, is one of the most well-known open source international repository of geographic data. General Transit Feed Specification, which provides transportation data to OpenTripPlanner, has been the standard for describing transit systems and platform for numerous applications. Together, when used to implement an instance of OpenTripPlanner, these software has been helping in traffic decongestion all over the world by assisting commuters to shift from using private transportation modes to public ones. Their potential however goes beyond providing multimodal public transportation routes. This paper aims to first discuss the researchers' experience in implementing a public transportation route planner for the purpose of traffic decongestion. After, the researchers would examine the prospective of using the system for disaster preparedness and recovery and concrete ways on how to realize them.

I. INTRODUCTION

Transportation has always been an integral part of human life. Everyday, people, animals, and products are always needed to be transported from one point to another. This puts mobility as one of the top most important need of modern life [1]. However, because of the continuous increase of human population and constant progression of human life, a lot of vehicles congest road networks [2]. Maximizing the use of roads networks can ease up this problem of network congestion. According to the National Household Travel Survey, private vehicle's average occupancy is only at 1.55 people [3], while jeepney's and busses are at 10.6 and 43.4 people accordingly [4]. These statistics show that to make efficient use of road networks, and to lessen the congestion along them, the number of private vehicles traversing main roads should be lessen. This reason drives governments from all over the world to push for a modal shift – urging motorists to change their transportation preference from using private transportation modes to public ones. This modal shift would respectively call for a shift from monomodal travelling –the use of only one transportation mode from origin to destination point – to a multimodal one – the use of more than one transportation mode. This is because of the obvious reason that each transportation mode and agency has its corresponding servicing area and function in the big picture of the transportation system. One concrete way to help the government push for this modal shift is by providing commuters with relevant and reliable travel information as it has been discovered that lack of public transportation information increases anxiety level and apprehension of commuters that holds them back from using public transportation modes [2].

The push for modal shift became the reason for the popularization of "Route Planners" – systems that calculate the best path from the user's inputted origin and destination points. The forefather of this kind of system only handled monomodal travelling, is mainly focused on catering to private vehicle driving, and mostly simply returns the shortest path between the given points. Current versions of route planners are now running different public transportation modes and are able to integrate them to produce a consolidated route, these systems are now more widely known as "Public Transportation Route Planners." Most of these route planners also now have the capability to take into consideration many objective functions to compute for the "best" path; this includes cheapest travel cost, shortest travel distance, and shortest travel time among others [5]. Many developers believe that these kinds of system were meant to offer help different kinds of commuters by providing information to, and to help push for the modal shift. The researchers also believe that an implementation of a public transportation route planner running on the road networks of Metro Manila would help ease the ever-worsening traffic in the city. They however see more potential with these systems, as they can also help in disaster management.

The researchers aim to share their knowledge about OpenTripPlanner, OpenStreetMap, and General Transit Feed

Specification, their own experience in implementing OpenTripPlanner with the transportation network of Metro Manila, and the potential of the produced system in helping in disaster relief and recovery. The produced system would compute for an array of suggested multimodal public transportation routes from the user's inputted origin and destination points. The suggested routes would be supplemented by useful travel information that the researchers see as important for most commuters, such as estimated travel cost, estimated travel distance, and estimated travel time. Going beyond the produced system's usual usage of providing multimodal public transportation routes, it can also be used to aid in disaster relief by computing for the best route trucks or tanks should follow when rescuing, distributing relief goods and supplies, and relocating survivors; it can help in disaster recovery by helping in logistics planning for rebuilding infrastructures.

## II. LITERATURE REVIEW

Public transportation route planners' popularity has been continually increasing all over the world. They have been implemented in different countries to aid different kinds of commuters with their travel. Although they follow different frameworks and methodologies, they all meet in their final goal of providing multimodal public transportation routes to entice commuters to use public transportation modes over private vehicles. This section would discuss some distinguished examples of this type of system used for its main value only.

### A. International Context

In the Eindhoven region (the Netherlands), implemented a Dijkstra-based Advanced Traveler Information System (ATIS) that ran on five (5) transportation modes [6]. In Vienna Austria, implemented a small-scaled application called "the Personal Travel Companion. [7]" In the US, Z. R. developed a distributed trip planning system that integrates different trip planning system in the country [8]. Another similar system, called "Smart Traveller Information Service" (STIS), was produced by S. Brennan and his team in Dublin Ireland [9]. In Tehran, R. Abbaspour and company created a multimodal public transportation route planner based on a genetic algorithm [10]. In Hong Kong, an Internet-based comprehensive public transport enquiry system (PTES) was developed by L. Pu-Cheng [11]. As a capstone for their research, F. H. Meng's group developed a system called "Route Advisory System" (RADS) that run the algorithm they created and implemented it with the public transportation of Singapore [12]. V. Spitadakis's group implemented WISETRIP in Greece, an international multimodal journey planning and delivery of personalized trip information [13]. Also developed in Greece, by K. Zografos and company is ENOSIS, a passenger information and trip planning system that aims to provide both urban and interurban multimodal planning services with real-time travel information [14]. Some other notable route planners include researches of L. Zhang et al. [15], J. Q. Li [16], and J. Jariyasunan's group. [17] who all created systems to support mobile-based advanced traveller information systems in the context of California. J. M. Su and company developed a multimodal trip planning system for intercity transportation in Taiwan using a search algorithm that considers the transit network, timetable, and the restrictions on access stops [18]. P Kumar *et al.* developed a GIS-based multimodal transportation system for Hyderabad city in India [19].

### B. Philippine Context

Google maps, which is perhaps one of the most well known route planning system in the world, has its own version of public transportation route planner called Google Transit. The system however only incorporates buses, MRT, LRT1, and LRT2 routes; jeepneys, which is the main transportation mode used in the Philippines, is not included in the system. Transit.com.ph, a system produced by the Philippine Transit App Challenge 2013, aimed to produce a local version of Google Transit by adding more local transportation modes. This system is however currently running the transportation network of Cebu city only; the routes of Metro Manila are not yet presently implemented. Thus, it is only able to return the walking route from the origin to the destination point. Sakay.ph, also produced by the Philippine Transit App Challenge 2013, was successful in creating a local version of the Google Transit application. Their system is able to compute for an array of routes for the given origin and destination point, complete with supplementary transportation information. They were also able to incorporate jeepney routes, which was the main shortcoming of Google Transit. However, it has been reported that some of its route suggestions are not the actual routes used by commuters. It is notable that the domain of public transportation route planners is helping commuters all around the world - even in the Philippines. There are however almost no studies and no published researches in the said field in the country.

Figure 1 is an illustration of the tree diagram representation of notable public transportation route planners. It summarizes the position of known public transportation route planners as compared to the proposed system in a graphical manner. As seen, PTES, RADS, WISETRIP, ENOSIS, Transit.com.ph, and Sakay.ph, are the closest systems to the researchers' proposal. Although these systems already implement most of the functionalities the researchers wish to put in their proposed route planner, none of them fully encapsulate the goals of the researchers. Most route planners only compute for the shortest path while the researchers want to give users all possible multimodal paths from the origin to the destination. Most route planners only take into consideration the shortest travel time, the proposed system would return an array of possible routes so commuters would have a number of options to choose from. Some planners only return the computed path, the proposed system would provide additional travel information such as estimated travel time, transportation expense, and distance to be travelled. Since other systems are in the business for money, they do not keep user information for other purposes, however the proposed system would save all user information and past sessions as one of its end goal is to produce analytical studies for the betterment of the transportation sector in the Philippines.

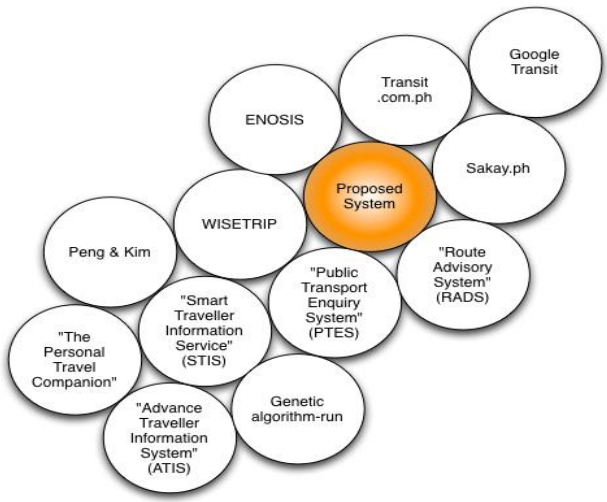

Figure 1 Public Transport Route Planners Tree Diagram

### III. FRAMEWORK

As stated in the introduction, the researchers implemented an instance of OpenTripPlanner with the help of OpenStreetMap and General Transit Feed Specification. This section would discuss the framework behind the research and system implementation.

#### A. OpenTripPlanner (OTP)

OpenTripPlanner is an open source, multimodal trip planning system collaboratively developed by a team of passionate developers from across the world, coordinated by OpenPlans and TriMet. It is a collaborative effort among TriMet, OpenPlans, and the developers of Five Points, OneBusAway, and Graphserver, and was identified as the most promising open source multimodal trip planning software system with an active developer community, as reported by B McHugh [20]. The system would be set up using biking and walking data from OSM and GTFS data from relevant transit agencies, both discussed in details below.

OpenTripPlanner contains most of the main features desired of in a multimodal public transportation trip planner. It is also important to note that it has an active community that is continuing to add new features and improve the software. What adds to the system's effectiveness and efficiency is its flexibility in being able to accommodate open data specifications and repositories, such as GTFS and OSM, to be able to support multimodal trip planning. It is also able to accept other sources of data including USGS National Elevation Dataset (NED) files and ESRI's shape files to help supplement OSM data, as stated in S. J. Barbeau's technical report [21].

Also according to S. J. Barbeau's final report, the next step to extend research of the OpenTripPlanner project is to create a public pilot deployment of the said system for a city or a country. A pilot deployment of OpenTripPlanner would be very valuable to researchers as a base for conducting studies on topics such as including applying changes to the OSM tagging system to improve its usefulness for multimodal trip planners, research on what information needs to be communicated to travellers for multimodal trips and how best to effectively communicate them, and much more. Congruently, a working example of the system would help stir up interest by other communities, not only in OpenTripPlanner but also in similar trip planning software [21].

#### B. OpenStreetMap (OSM)

OpenStreetMap is maintained by the non-profit OpenStreetMap foundation. It is an open sourced and freely available international repository of geographic data that individuals contribute to about their local community. The general public and organizations such as transit agencies with public-domain data add geographic information into the system. Although most contributors are volunteers, there have been an increasing number of both commercial organization and government bodies contributing to the project. These data contributors gather information by driving, cycling, or walking along streets and paths, recording their every move using Global Positioning System (GPS). Data from the system can easily be shared and viewed by requesting a download or viewing a simple webpage-based application. Anyone with an interest in providing and using data about any area is free to do so. Also, when someone finds an error in the data, he or she is free to correct it accordingly. Because the system uses crowd sourcing, the ability to gather information from multiple different sources, the general public is able to contribute large amounts of data instead of only one organization being held responsible for creating and updating data. There are a number of open-source software tools that exist to work well with OSM data. Software like them allow communities to collaborate on software projects, providing a robust and stable base for improvements, as also stated in S. J. Barbeau's technical report [21].

#### C. General Transit Feed Specification (GTFS)

Google's offer of a free online trip planner based on the General Transit Feed Specification has made GTFS in practice the standard for describing transit systems and platform for many other web and mobile applications. Google Transit trip-planning service was able to encourage 125 public transportation agencies in the U.S. alone to put their data into the GTFS format for Google Transit. However, although GTFS format is open, Google Transit does not serve as a common data repository where the underlying information can be freely shared without licensing fees or copyright restrictions, as also stated in S. J. Barbeau's technical report [21].

TABLE I.
REQUIRED GTFS FILES FOR OPENTRIPPLANNER

| Textfile | Textfile Information |
|---|---|

| Agency | One or more transit agencies that provide transportation data |
|---|---|
| Calendar | Date for service IDs using a weekly schedule |
| Frequencies | Headway (time between trips) for routes with variable frequency of service |
| Routes | Transit routes. A route is a group of trips that are displayed to riders as a single service |
| Shapes | Rules for drawing lines on a map to represent a transit organization routes |
| Stop Times | Times that a vehicle arrives and departs from individual stops for each trip |
| Stops | Individual location where vehicles pick up or drop off passengers |
| Trips | Trip for each route. A trip is a sequence of two or more stops that occur at a specific time |

Table 1 summarizes the required General Transit Feed Specification files OpenTripPlanner needs to be able to create the graph and compute for multimodal public transportation routes.

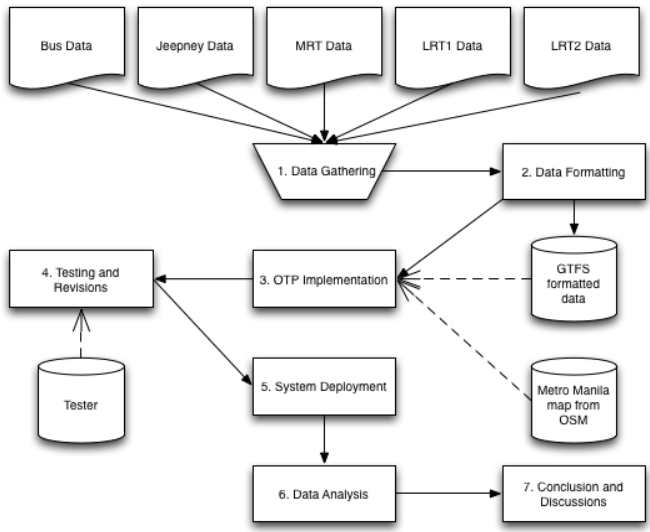

Figure 2. Framework

Figure 2 is a diagram summarizing the theoretical framework behind this research. Data was collected from LTFRB and other governing transportation agencies. Separately, DOTC also produced a set of data that is formatted according to the standards of General Transit Feed Specification (GTFS). Aside from being validated through crowd source survey, GTFS data from LTFRB were also personally validated by the researchers. GTFS data along with the extracted Metro Manila map from OSM would be inputted to OpenTripPlanner's graph builder module to produce the graph of Metro Manila. This graph of Metro Manila would be the basis of the system to produce the computed routes for a given origin and destination points. The computed routes would then be subjected to a tester to assess the system's correctness and effectiveness.

IV. RESULTS

This section is dedicated to discussing the significant details regarding the initial implementation of the OpenTripPlanner instance.

Before implementing the actual system, the graph of the subjected area with the use of OTP's graph builder module is first needed to be created. For this module to complete its task, two inputs are required: the OSM data of the area and the GTFS data of the area's transportation system. The home website of OSM does not allow direct extraction of large amounts of data so a third party website, Metro Extracts (http://metro.teczno.com/), was used to download the map of Metro Manila. The Department of Transportation and Communication's Public Transit Information Service website's latest published GTFS data was directly downloaded.

After creating the graph, OTP's web API is needed to be configured to properly run the web application. Although the basic set-up of the application is already able to calculate for multimodal public transportation routes, some modifications were made to make the application user-friendly. These modifications includes setting the origin and destination input textboxes visible so that users may opt not to click on the map provided, and the utilization of the geocoder module so that users can input actual places instead of longitude and latitude values.

Testing of the application was done by randomly selecting popular places in Metro Manila and inputting them as origin and destination points in the provided textboxes, and randomly clicking on the provided map. Figure 3 is screenshot of the system's interface when used to plan a trip from Ateneo de Manila University to De La Salle University. For this specific instance, the system was able to compute for 3 different routes between the two points.

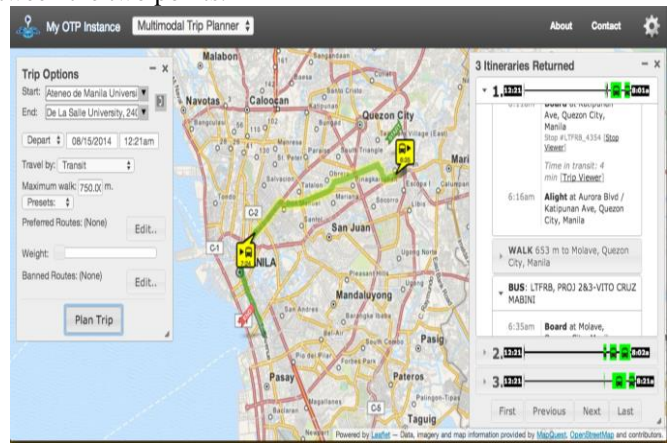

Figure 3 Route from Ateneo de Manila University to De La Salle University

## V. ANALYSIS

OpenTripPlanner is always able to compute for a multimodal public transportation route from the given origin to the destination, as proven by the testing done by the researchers. In events when one point goes beyond the boundaries of the subjected area, OTP computes for a route from the validly inputted point to a valid endpoint nearest the invalidly inputted point. Walking routes were added to connect the valid endpoint and the invalidly inputted point. In occasions where both origin and destination points lie beyond the boundaries of Metro Manila, an error message returns to the user stating "Trip is not possible. You might be trying to plan a trip outside the map boundary."

## VI. CONCLUSIONS

The researchers implemented a public transportation route planner using OpenTripPlanner, with the help of OpenStreetMap and GTFS data. The latest published GTFS data from LTFRB was used, but as it is continuously being completed and updated, the system should be updated with every new GTFS data publication.

The difficulty in the implementation of a route planner lies in data gathering, formatting, and updating. These are the focus of the next steps of the research.

As mention in the first part of this paper, the possibilities of OpenTripPlanner goes beyond its basic value of providing multimodal public transportation routes. Its routing capabilities can be used for the purpose of disaster relief and recovery. When a disaster strikes a certain region, the OSM map of the affected area would be downloaded and used for the system. Some modifications to the map can be done like destroying roads and buildings which were destroyed by the recent disaster. GTFS data should be created, following the details on table I. GTFS data need not be detailed as only servicing trucks and tanks would need route calculation for purposes of relied in rescuing, distributing relief goods and supplies, and relocation; disaster recovery by helping in logistics planning for rebuilding infrastructures.

## VII. ACKNOWLEDGMENT

This research is supported by the Department of Science and Technology - Engineering Research and Development for Technology (DOST-ERDT) and Philippine Higher Education Research Network (PHERNET).

REFERENCES

[1] S. M. Kumari and N. Geethanjali, A Survey on Shortest Path Routing Algorithms for Public Transport Travel, Global Journal of Computer Science and Technology, 9 (5): 73 - 75
[2] S. Kenyon and G. Lyons The value of integrated multimodal traeller information and its potential contribution to modal change. Transportation Research Part F: Traffic Psychology and Behaviour, 6(1):1–21, 2003.
[3] US Department of Transportation Federal Highway Admission, 2000 NHTS Average Vehicle, 2009
[4] World Bank, Implementation Completion and Results Report on a Loan in the Amount of $60.0 Million and a Global Environmental Facility Grant in the Amount of SDR 1.0 Million (US $ 1.3 Equivalent) to the Republic of the Philippines for the Metro Manila Urban Transport Integration Project, June 2011.
[5] F. H. Meng, Y. Lao, L. H. Wai, and L. H. Chuin, A multi- criteria, multi-modal passenger route advisory system. National Science and Technology Board (NSTB) Singapore, 1999
[6] J. Zhang, Arentze, A. Theo, and H. Timmermans, A multi- modal transport network model for advanced traveler information system. JUSPN, 4(1):21–27, 2012.
[7] K. Rehrl, N. Goll, S. Leitinger, and S. Bruntsch, Combined indoor/outdoor smartphone navigation for public transport travellers. In Proceedings of the 3rd Symposium on LBS & TeleCartography, volume 2005, pages 235– 239, 2005.
[8] Z. R. Peng and E. Kim, A standard-based integration framework for distributed transit trip planning systems, Journal of Intelligent Transportation Systems, 12(1):13–28, 2008
[9] S. Brennan and R. Meier, STIS: Smart travel planning across multi- ple modes of transportation. In Intelligent Transportation Systems Conference, 2007. ITSC 2007. IEEE, pages 666–671, IEEE, 2007
[10] Abbaspour, R. A. and Samadzadegan, F. (2010) An evolutionary solution for multimodal shortest path problem in metropolises. Computer Science and Information Systems, 7(4):789–811.
[11] Pun-Cheng, L. (2012) An interactive web-based public transport enquiry system with realtime optimal route computation. Intelligent Transportation Systems, IEEE Transactions, 13(2):983–988.
[12] Meng, F. H., Lao, Y., Wai, L. H., and Chuin, L. H. (1999) A multi- criteria, multi-modal passenger route advisory system. National Science and Technology Board (NSTB) Singapore.
[13] Spitadakis, V., and Fostieri, M. (2012) Wisetrip-international multimodal journey planning and delivery of personalized trip information. Procedia-Social and Behavioral Sciences, 48:1294–1303.
[14] Zografos, K. G., Androutsopoulos, K. N., and Vassilis Spitadakis. (2009) Design and assessment of an online passenger information system for integrated multimodal trip planning. Intelligent Transportation Systems, IEEE Transactions, 10(2):311–323.
[15] Zhang, L., Li, J. Q., Zhou, K., Gupta, S. D., Li, M., Zhang, W.B., Miller, M. A., and Misener, J. A. (2011) Traveler information tool with integrated real-time transit information and multimodal trip planning. Transportation Research Record: Journal of the Transportation Research Board, 2215(1):1–10.
[16] Li, J. Q., Zhou, K., Zhang, L., and Zhang, W. B. (2010) A multimodal trip planning system incorporating the park-and-ride mode and real-time traffic/transit information. In Proceedings ITS World Congress, volume 25, pages 65–76.
[17] Jariyasunant, J., Kerkez, B., Sengupta, R., Glaser, S., and Bayen, A. (2011) Mobile transit trip planning with real-time data.
[18] Su, J. M., Chang, C. H., and Ho, W. C., (2008) Developmentoftripplanning systems on public transit in taiwan. In Networking, Sensing and Control, 2008. ICNSC 2008. IEEE International Conference, pages 791–795. IEEE.Zhang, J., Arentze, J. A., and Timmermans, H. (2012) A multimodal transport network model for advanced traveler information system. JUSPN, 4(1):21–27.
[19] Kumar, P., Singh, V., and Reddy, D. (2005) Advanced traveler information system for hyderabad city. Intelligent Transportation Systems, IEEE Transactions, 6(1):26–37.
[20] McHugh, B. (2011) The opentripplanner. Metro 2009-2011 regional travel operations grant final report, TriMet, August 2011.
[21] Barbeau, S. J., and Hillsman, E. L. (2011) Enabling cost-effective multimodal trip planners through open transit data. Technical

report, University of South Florida (USF) National Center for Transit Research, May 2011.